\documentclass[fleqn,usenatbib]{mnras}

\usepackage{newtxtext,newtxmath}

\usepackage[T1]{fontenc}

\DeclareRobustCommand{\VAN}[3]{#2}
\let\VANthebibliography\thebibliography
\def\thebibliography{\DeclareRobustCommand{\VAN}[3]{##3}\VANthebibliography}

\usepackage[normalem]{ulem}

\usepackage{graphicx}	
\usepackage{bbm}    

\usepackage{xcolor}
\usepackage{xspace}

\newcommand{\hyrange}{~--~}
\newcommand{\gaia}{\textit{Gaia}\xspace}
\newcommand{\code}[1]{\textsc{#1}\xspace} 
\newcommand{\acknowledgeSoftware}[1]{\code{#1} \citep{#1}\xspace}

\newcommand{\masPerYr}{$\rm mas\,yr^{-1}$\xspace}
\newcommand{\kmPerSec}{$\rm km\,s^{-1}$\xspace}

\newcommand{\G}{{\rm G}}
\newcommand{\SN}{S_{\rm N}}
\newcommand{\Mv}{M_{\rm V}}
\newcommand{\Ngc}{N_{\rm gc}}
\newcommand{\Lgal}{L_{\rm gal}}
\newcommand{\MvLim}{\Mv^{\rm lim}}
\newcommand{\SigmaIn}{\Sigma_{\rm in}}
\newcommand{\SigmaOut}{\Sigma_{\rm out}}
\newcommand{\SigmaInLim}{\SigmaIn^{\rm lim}}
\newcommand{\SigmaInObs}{\SigmaIn^{\rm obs}}
\newcommand{\Gaussian}[2]{\mathcal{N}({#1},\,{#2}^2)}

\title[Searching for globular clusters using \gaia DR2]{Search for globular clusters associated with the Milky Way dwarf galaxies using \textit{\textbf{Gaia}} DR2}

\author[K.-W. Huang et al.]{
Kuan-Wei Huang$^{1}$\thanks{E-mail: kuanweih@andrew.cmu.edu} and Sergey E. Koposov$^{2,1,3}$
\\
$^{1}$McWilliams Center for Cosmology, Dept. of Physics, Carnegie Mellon University, Pittsburgh, PA, 15213, USA\\
$^{2}$Institute for Astronomy, University of Edinburgh, Royal Observatory, Blackford Hill, Edinburgh EH9 3HJ, UK\\
$^3$Institute of Astronomy, Madingley Rd, Cambridge, CB3 0HA
}

\date{Accepted XXX. Received YYY; in original form ZZZ}

\pubyear{2020}

\begin{document}
\label{firstpage}
\pagerange{\pageref{firstpage}--\pageref{lastpage}}
\maketitle

\begin{abstract}
We report the result of searching for globular clusters (GCs) around 55 Milky Way satellite dwarf galaxies within the distance of 450~kpc from the Galactic Center except for the Large and Small Magellanic Clouds and the Sagittarius dwarf. 
For each dwarf, we analyze the stellar distribution of sources in \gaia DR2, selected by magnitude, proper motion, and source morphology. 
Using the kernel density estimation of stellar number counts, we identify eleven possible GC candidates. 
Crossed-matched with existing imaging data, all eleven objects are known either GCs or galaxies and only Fornax GC 1\hyrange6 among them are associated with the targeted dwarf galaxy. 
Using simulated GCs, we calculate the GC detection limit $\MvLim$ that spans the range from $\MvLim \sim -7$ for distant dwarfs to $\MvLim \sim 0$ for nearby systems. 
Assuming a Gaussian GC luminosity function, we compute that the completeness of the GC search is above 90 percent for most dwarf galaxies. 
We construct the 90 percent credible intervals/upper limits on the GC specific frequency $\SN$ of the MW dwarf galaxies: $12 < \SN < 47$ for Fornax, $\SN < 20$ for the dwarfs with $-12 < \Mv < -10$, $\SN < 30$ for the dwarfs with $-10 < \Mv < -7$, and $\SN < 90$ for the dwarfs with $\Mv > -7$. 
Based on $\SN$, we derive the probability of galaxies hosting GCs given their luminosity, finding that the probability of galaxies fainter than $\Mv = -9$ to host GCs is lower than 0.1.
\end{abstract}



\section{Introduction}
\label{sec:intro}

Globular clusters (GCs) are some of the oldest luminous observable objects with ages comparable to the age of the Universe \citep{vandenberg2013}. 
Characterized by being compact and bright, GCs typically have masses of $10^4$\hyrange$10^6 M_{\sun}$, luminosities of $\Mv = -5$ to $-10$, and sizes of a few parsecs \citep{1991ARA&A..29..543H, doi:10.1146/annurev.astro.44.051905.092441}. 
GCs might have played an important role in the early formation of galaxies, and they could have been the potential drivers of cosmic reionization \citep{BoylanKolchin2018} despite the issues with the escape fraction of ionizing radiation \citep{2018MNRAS.475.3121H,2020MNRAS.492.4858H}.  
However, the formation of GCs themselves remains an open question in astrophysics \citep[some recent literature that discuss the formation of GCs, e.g.][]{2018NatAs...2..725H,2019MNRAS.486.5838R,2019MNRAS.486..331C,2019MNRAS.482.4528E,2020MNRAS.493.4315M}. 
For detailed reviews of GCs, we refer readers to \citet{2004ARA&A..42..385G}, \citet{doi:10.1146/annurev.astro.44.051905.092441}, and \citet{2019A&ARv..27....8G}.

In the Milky Way (MW), the number of known GCs has increased to around 150 \citep{1996AJ....112.1487H, 2010arXiv1012.3224H} since the first one was discovered in 1665 by Abraham Ihle. 
While some of these GCs that are more concentrated around the Galactic Center are believed to have been formed {\it in-situ} \citep{1997AJ....113.1652F, 1999AJ....117..855H}, the ones in the outskirts are believed to have been accreted together with their parent dwarf galaxies \citep[e.g.][]{1978ApJ...225..357S, Mackey2004, 2018Natur.555..483B, Kruijssen2019}, which were destroyed by tides. 
Some of the GCs however can still be found within the MW satellites themselves, offering a window on the formation of GCs in dwarf galaxies. 
The three most luminous MW satellites, the Large and Small Magellanic Clouds (LMC and SMC) and the Sagittarius dwarf spheroidal galaxy, have large populations of GCs \citep{2003MNRAS.338...85M, 2003MNRAS.338..120M, 2003MNRAS.340..175M, 2005ApJS..161..304M}. 
In particular, the clusters of the Sagittarius dwarf are spread out along the stellar stream \citep{1995MNRAS.275..429L, 2003AJ....125..188B, 2017MNRAS.468...97L, Vasiliev2019}, and the SMC has a large population of star clusters in general but few of them are classically old GCs. 
The only other two MW satellite galaxies known to possess GCs are the Fornax dwarf spheroidal galaxy which is the fourth most luminous MW satellite with six GCs, and the Eridanus 2, an ultra-faint system containing a faint cluster \citep{Koposov_2015, 2016ApJ...824L..14C}.


The fact that some GCs in the MW still have been found until recently \citep{Koposov_2015, 2017MNRAS.470.2702K,2019ApJ...875L..13W} motivates us to further search for possibly missing ones. 
Intuitively, faint GCs within dwarf galaxies are more likely to have been missed, especially when located within luminous dwarf galaxies where the ground-based data can be crowded e.g. Fornax~6. 
Instead of looking for this kind of objects by chance, we apply the systemic overdensity searching algorithm (which will be explained in Section~\ref{sec:method}) to the areas around the MW satellite galaxies within the distance of 450~kpc except for the three most luminous ones: the LMC, the SMC, and the Sagittarius dwarf. 
That is, we target the areas where GCs are likely to lurk from previous inspections of deep imaging to look for overdensities in dense dwarfs.

Focusing on a small area of the sky, a targeted search is less computationally expensive so that it can afford a lower detection threshold. 
For each targeted area, we investigate the stellar distribution in the \gaia data to detect possible GC candidates (see Section~\ref{subsec:gaia} for more detail about \gaia and the dataset).
Thanks to the high angular resolution that exceeds most ground-based surveys, \gaia allows us to detect previously missed objects that are not well resolved or missed by ground-based searches.
For instance, \citet{2017MNRAS.470.2702K} has found star clusters in \gaia that were missed by previous searches.

We organize the paper as follows. 
In Section~\ref{sec:method}, we explain the methodology with more detail about the \gaia data, sample selection, and kernel density estimation procedure.
In Section~\ref{sec:result}, we demonstrate the main results of the detection. 
In Section~\ref{sec:discussion}, we discuss the limit and completeness of the detection, the inferred specific frequency of GCs, and the derived probability of dwarfs to host GCs based on our findings. 
In Section~\ref{sec:conclusion}, we conclude the paper.

\section{Methodology}
\label{sec:method}

\begin{figure*}
    \includegraphics[width=2\columnwidth]{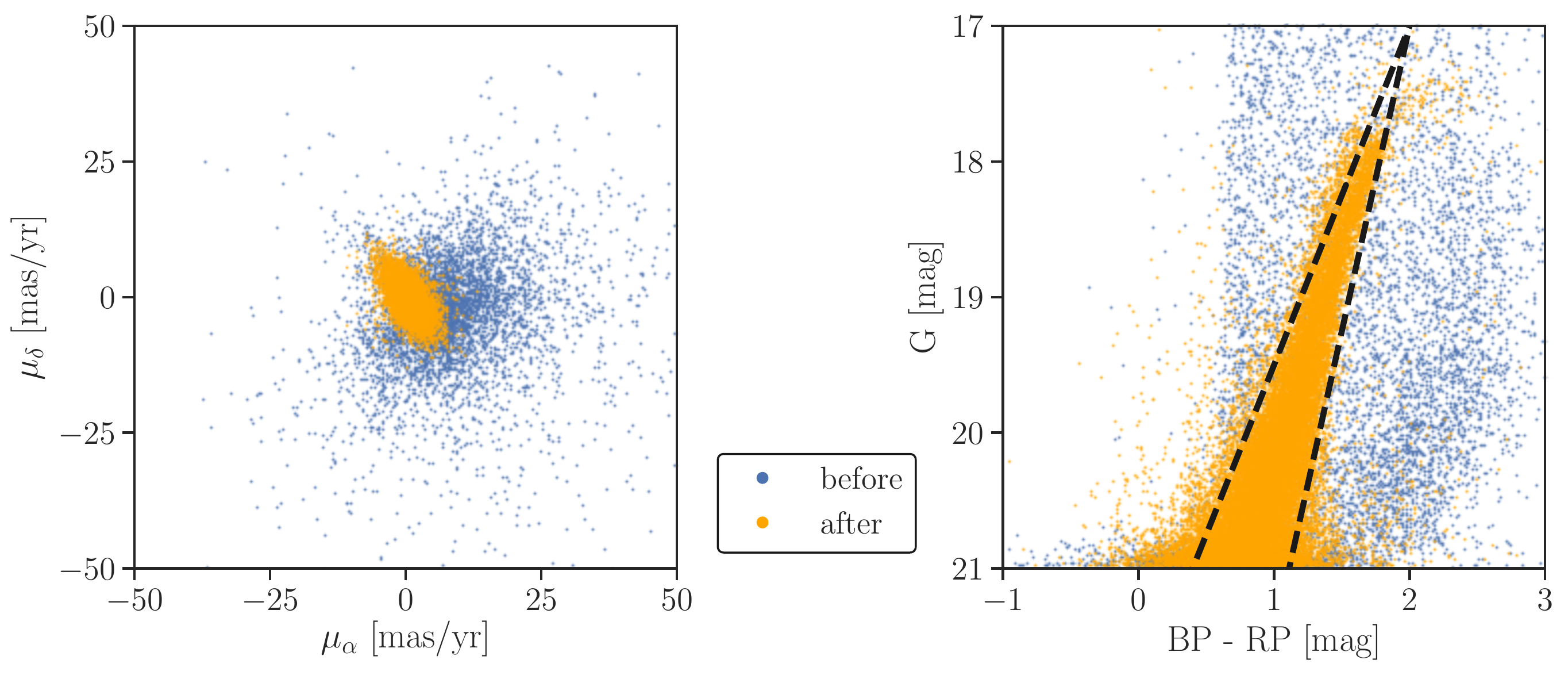}
    \caption{The \gaia sources around the Fornax dwarf before (blue) and after (orange) the proper motion selection defined in Equation~\ref{eq:pmcut}. \textbf{Left}: the distribution in proper motion space. \textbf{Right}: the color-magnitude diagram. The black dashed lines define a lasso to roughly distinguish possible member stars in the red-giant branch of Fornax.} 
    \label{fig:method_ex_Fornax_propermotion}
\end{figure*}

\begin{figure*}
    \includegraphics[width=2\columnwidth]{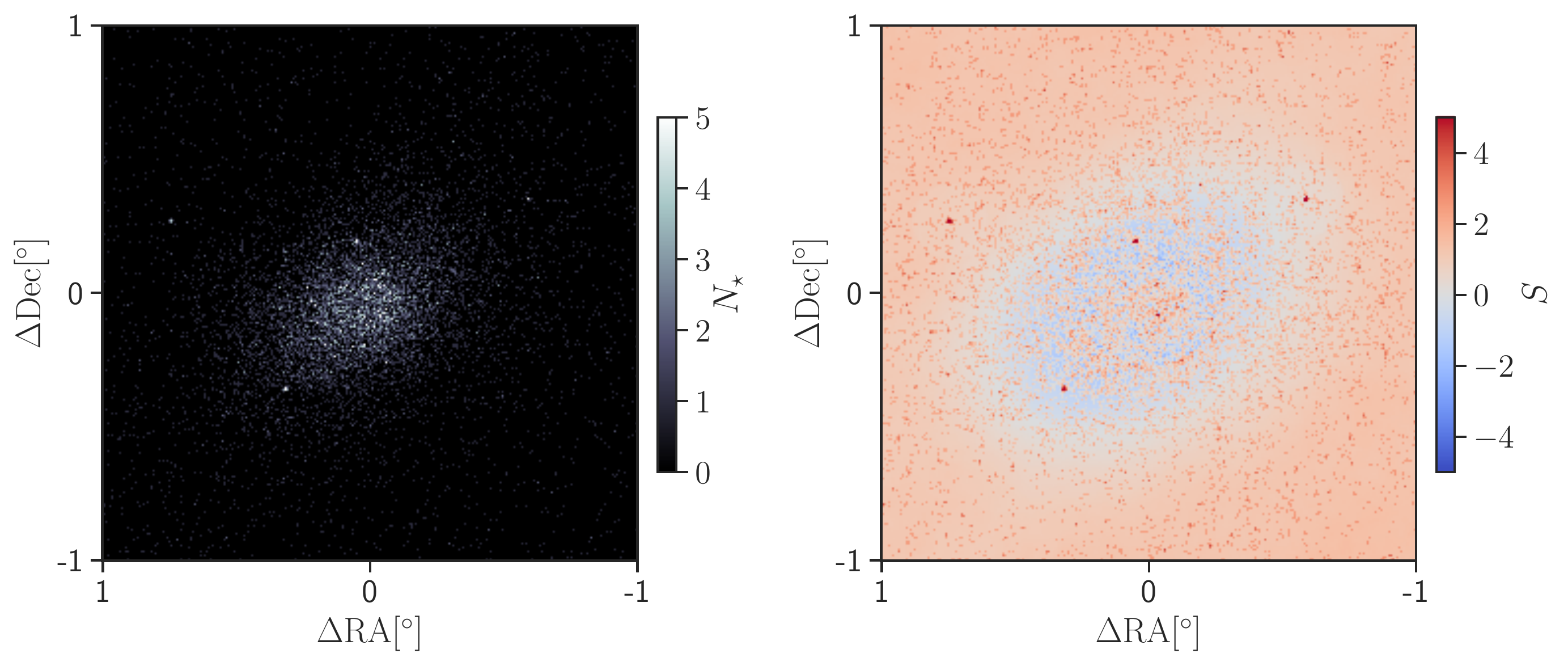}
    \caption{\textbf{Left}: the two-dimensional histogram of \gaia DR2 sources selected using Equations~\ref{eq:g_cuts}, \ref{eq:astro_noise_cuts}, and \ref{eq:pmcut} around the Fornax dwarf. \textbf{Right}: the over-density significance ($S$) map according to Equation~\ref{eq:calc_significance}.} 
    \label{fig:method_ex_Fornax}
\end{figure*}

\subsection{\textit{\textbf{Gaia}} DR2 and data selection}
\label{subsec:gaia}
The space-based astrometric mission \gaia was launched by the European Space Agency in 2013 and started the whole-sky survey in 2014 \citep{2016A&A...595A...1G}. 
Released in 2018, the second \gaia data release (\gaia DR2) contains the data collected during the first 22 months of the mission \citep{2018A&A...616A...1G} and has approximately 1.7 billion sources with 1.3 billion parallaxes and proper motions. 
\gaia DR2, therefore, provides high-resolution stellar distribution in the MW for us to look for possibly missing GCs around the MW dwarf galaxies. 
The overall scientific validation of the data is described in \citet{2018A&A...616A..17A}. 

The entire analysis of this paper utilizes the \texttt{GAIA\_SOURCE} catalog of \gaia DR2 \citep{DocumentationRelease1.2}, particularly the position \texttt{ra} and \texttt{dec} ($\alpha$ and $\delta$), the proper motion (PM) \texttt{pmra} and \texttt{pmdec} ($\mu_{\alpha}$ and $\mu_{\delta}$), the G-band magnitude \texttt{phot\_g\_mean\_mag} (G), and the value of the \texttt{astrometric\_excess\_noise} parameter ($\epsilon$).  
\citet{2018A&A...616A...1G} contains the detail on the contents and the properties of this catalog. 
We use this dataset to identify stellar density peaks as possible candidates of GCs around in the vicinity and inside nearby dwarf galaxies.

Throughout the whole paper, we apply two main selection cuts on the \gaia catalog. 
The first selection is 
\begin{equation}
\label{eq:g_cuts}
    17 < \G < 21. 
\end{equation}
The faint-magnitude cut $\G < 21$ approximately corresponds to the faint-end limit of \gaia DR2; \citet{2018A&A...616A...1G} reported that only 4 percent of the sources are fainter than $\G = 21$ and those sources lack PMs and parameters. 
There are two reasons for the bright-magnitude cut $\G > 17$. 
The first reason to get rid of the bright stars is that the foreground contamination dominates at bright magnitudes. 
Conversely, the expected rapid rise of the stellar luminosity function for the majority of GCs and dwarf galaxies at reasonable distances from the Sun at $\G > 17$ results in the majority of stars being fainter than $\G = 17$. 
The other reason is that most bright GCs with large numbers of $\G < 17$ stars would have likely been detected already. 
The second selection criterion is
\begin{equation}
    \label{eq:astro_noise_cuts}
    \ln \epsilon < 1.5 + 0.3 \max \{\G - 18, 0\}.
\end{equation}
This cut is used to reject potentially extended sources \citep[see][for more detail]{2017MNRAS.470.2702K,2019ApJ...875L..13W}.

Another optional selection that we use to further clean the source list is based on the PM, with the goal of removing sources whose PMs are different from the mean PM of a given targeted dwarf galaxy, as these sources are less likely to be member stars of the given dwarf. 
For each targeted dwarf, we exclude stars with PMs ($\mu_{\alpha}$, $\mu_{\delta}$) differing from a systemic PM of the dwarf ($\mu_{\alpha}^{\rm dwarf}$, $\mu_{\delta}^{\rm dwarf}$) by more than three times the PM uncertainty ($\sigma_{\mu_{\alpha}}$, $\sigma_{\mu_{\delta}}$). 
That is, only the stars satisfying
\begin{equation}
\label{eq:pmcut}
    \sqrt{ (\mu_{\alpha} - \mu_{\alpha}^{\rm dwarf})^2 
         + (\mu_{\delta} - \mu_{\delta}^{\rm dwarf})^2} 
         < 3 \sqrt{\sigma_{\mu_{\alpha}}^2 + \sigma_{\mu_{\delta}}^2}
\end{equation}
survive after the PM selection.

For example, Figure~\ref{fig:method_ex_Fornax_propermotion} shows the \gaia sources around the Fornax dwarf before and after the PM selection in Equation~\ref{eq:pmcut}.  
The source distribution in PM space in the left panel shows that there are many foreground sources with PMs that are 10\hyrange100~\masPerYr different from the PM of the dwarf. 
This PM selection is thus applied to remove this kind of contamination; the sources colored in orange survive after the selection. 
It is worth noting that the PM uncertainty of the studied dwarfs is around the order of $10^3$\hyrange$10^5$~\kmPerSec (see Table~\ref{tab:target-dwarfs}) which is much larger than the typical velocity dispersion of dwarf galaxies around the order of 10~\kmPerSec \citep{2007ApJ...667L..53W} or 0.02~\masPerYr if at 100~kpc, so the survived sources under this PM selection still have a fairly large range of internal space velocity. 
To investigate the PM selection for the stars that are more likely to be member stars of the Fornax dwarf, we draw a lasso with the black dashed lines to roughly distinguish the member stars in the red-giant branch of Fornax from the other stars in the color-magnitude diagram in the right panel. 
For the stars that are likely to be member stars inside of the lasso, 91~percent of the sources survive after the PM selection, whereas most of the sources outside of the lasso are excluded. 
Moreover, in the left panel of Figure~\ref{fig:method_ex_Fornax}, the stellar distribution after the PM selection retains the shape of the Fornax dwarf. 

\subsection{Kernel density estimation}
\label{subsec:kde}

Convolving the spatial distribution of the data with various kernels is a common approach to identify the excess number of stars associated with a satellite or clusters in imaging data. 
The density is calculated by convolving all the data points interpreted as delta functions with different kernels, e.g. a moving average in \citet{2009AJ....137..450W}, two circular indicator functions in \citet{2019MNRAS.484.2181T} and Gaussian kernels in \citet{2008A&A...486..771K}, \citet{2008ApJ...686..279K}, and \citet{2015ApJ...813..109D}. 

To identify star clusters in dwarf galaxies, we use the kernel density estimation on the stellar distribution, while assuming the Poisson distribution of stellar number counts.
\begin{enumerate}
  \item We obtain the distribution of stars
        \begin{equation}
        \label{eq:dist_map}
            \Sigma(x, y) = \sum_i \delta(x - x_i, y - y_i)
        \end{equation}
        where $(x_i, y_i)$ is the position of the $i^{\rm th}$ star on the local coordinates which takes care of the projection effect.\footnote{
            In the algorithm, we always divide a targeted area into small patches with a side of $0.5^\circ$. For each patch centered at $(\alpha_0, \delta_0)$, we define the local coordinates $(x, y)$ with the origin of $(x_0, y_0) = (\alpha_0, \delta_0)$. Since the patch is very small, we approximate the projection effect as $x \approx (\alpha - \alpha_0) \cos \delta_0$ and $y = \delta - \delta_0$.
        }
  \item Using the circular indicator function with a given radius $R$ defined as 
        $\mathbbm{1} \left( x, y; R \right) = \left\{\begin{matrix}
        \ 1 \ {\rm if} \ x^2 + y^2 \leq R^2 \\ 
        \ 0 \ {\rm otherwise} \ \ \ \ \ \ \ \ \ \ 
        \end{matrix}\right.$, we define the inner kernel $K_{\rm in} (x, y; \sigma_1) = \mathbbm{1} \left( x, y; \sigma_1 \right)$, where $\sigma_1$ corresponds to the scale of GCs which is 3, 5, or 10~pc. 
        We then convolve $\Sigma(x, y)$ with $K_{\rm in} (x, y; \sigma_1)$ to estimate the number density of stars on the scale of $\sigma_1$ as 
        \begin{equation}
        \label{eq:n_inner}
            \SigmaIn (x, y) = \Sigma(x, y) * K_{\rm in} (x, y; \sigma_1).
        \end{equation}
  \item Defining the outer kernel 
        $K_{\rm out} (x, y; \sigma_1, \sigma_2) = \mathbbm{1} \left( x, y; \sigma_2 \right) - \mathbbm{1} \left( x, y; 2\sigma_1 \right)$, we convolve $\Sigma(x, y)$ with $K_{\rm out} (x, y; \sigma_1, \sigma_2)$ as 
        \begin{equation}
        \label{eq:n_outer}
            \SigmaOut (x, y) = \Sigma(x, y) * K_{\rm out} (x, y; \sigma_1, \sigma_2)
        \end{equation}
        to estimate the number density of stars on the annular area of radius between $2\sigma_1$ and $\sigma_2$, where $\sigma_2 > 2 \sigma_1$ and $\sigma_2$ corresponds to either the angular scale of parent dwarf galaxy or a fixed angular scale of $0.5^\circ$ (see more detail in the next paragraph). 
  \item We estimate the expected background number density within the inner kernel from 
        $\SigmaOut (x, y)$ through the ratio of the inner and outer areas
        \begin{equation}
        \label{eq:calc_lambda}
            \Sigma_{\rm bg} (x, y) = \frac{ \sigma_1^2 }{\sigma_2^2 - (2\sigma_1)^2} \, \SigmaOut (x, y). 
        \end{equation}
  \item We convert the tail probability of Poisson into the z-score of 
        the standard normal distribution to evaluate the significance as
        \begin{equation}
        \label{eq:calc_significance}
            S(x, y) =  F_{\rm N \left( 0,1 \right) }^{-1} \left( F_{{\rm Poi} \left( \Sigma_{\rm bg} \left( x, y \right) \right) } \left( \SigmaIn \left( x, y \right) \right) \right),
        \end{equation}
        where $F$ is the cumulative distribution function. 
\end{enumerate}
As an example, Figure~\ref{fig:method_ex_Fornax} shows the original two-dimensional histogram of the sources around the Fornax dwarf in the left panel and the significance map of that stellar distribution in the right panel. 
According to the significance map, we identify positive detection with significance higher than a certain significance threshold.
For nearby pixels with significance higher than the threshold, we merge them as one single positive detection if the radial distance between the pixels is shorter than the size of the inner kernel. 
We assign the maximum significance on the merged pixels as the detected significance and use the center of mass coordinates of the merged pixels as the detected position.

The main reason for $\sigma_2$ in step (iii) corresponding to either the angular scale of parent dwarf galaxy or the fixed angular scale of $0.5^\circ$ is that the kernel density estimates are biased in crowded areas, which may lead to missing objects around big dwarfs. 
Given a dwarf with a half-light radius of $r_h$, $\sigma_2$ is chosen to be $0.5 r_h$ or $0.5^\circ$ for pixels inside ($r < r_h$) or outside ($r > r_h$) of the dwarf respectively, where $r$ is the distance from the position of any pixel to the center of the dwarf. 
The latter large $\sigma_2$ of $0.5^\circ$ is to take care of the sparse outskirts of the dwarf. 
Besides, when dealing with the pixels outside of the dwarf, we exclude the effect of the pixels inside of the dwarf ($r_h < r < r_h + 0.5^\circ$) because the relatively high number density of stars in the dwarf will lead to over-estimate of $\SigmaOut (x, y)$ which will suppress the background estimate too much later.

\section{Results}
\label{sec:result}

\begin{table}
\caption{The nine known GCs and the two known galaxies found in our detection and their detected positions ($\alpha$ and $\delta$), significance values ($S$), and inner kernel sizes ($\sigma_1$). 
Only Fornax 1\hyrange6 are actual clusters belonging to their parent dwarf galaxy.}
\label{tab:known_gcs_found}
\begin{tabular}{lrrrrl}
      Objects & $\alpha$ [$^\circ$] & $\delta$ [$^\circ$]&    $S$ & $\sigma_1$ [pc] \\ 
\hline
      Fornax 1 &       40.5871 &      -34.1016 &  8.5 &              10 \\
      Fornax 2 &       39.6842 &      -34.8092 &  8.2 &              10 \\ 
      Fornax 3 &       39.9502 &      -34.2593 &  7.4 &              10 \\ 
      Fornax 4 &       40.0343 &      -34.5375 &   5.4 &              10 \\ 
      Fornax 5 &       39.2570 &      -34.1845 &   7.5 &              10 \\ 
      Fornax 6 &       40.0298 &      -34.4204 &   5.2 &              10 \\ 
     Palomar 3 &      151.3788 &        0.0731 &   7.3 &              10 \\ 
    Messier 75 &      301.5206 &      -21.9233 &  37.6 &              10 \\ 
      NGC 5466 &      211.3615 &       28.5321 &  37.7 &              10 \\ 
\hline
         Leo I &      152.1122 &       12.3001 &  37.2 &              10 \\ 
     Sextans A &      151.3799 &        0.0714 &   6.3 &              10 \\ 
\hline
\end{tabular}
\end{table}

\begin{figure*}
    \includegraphics[width=2\columnwidth]{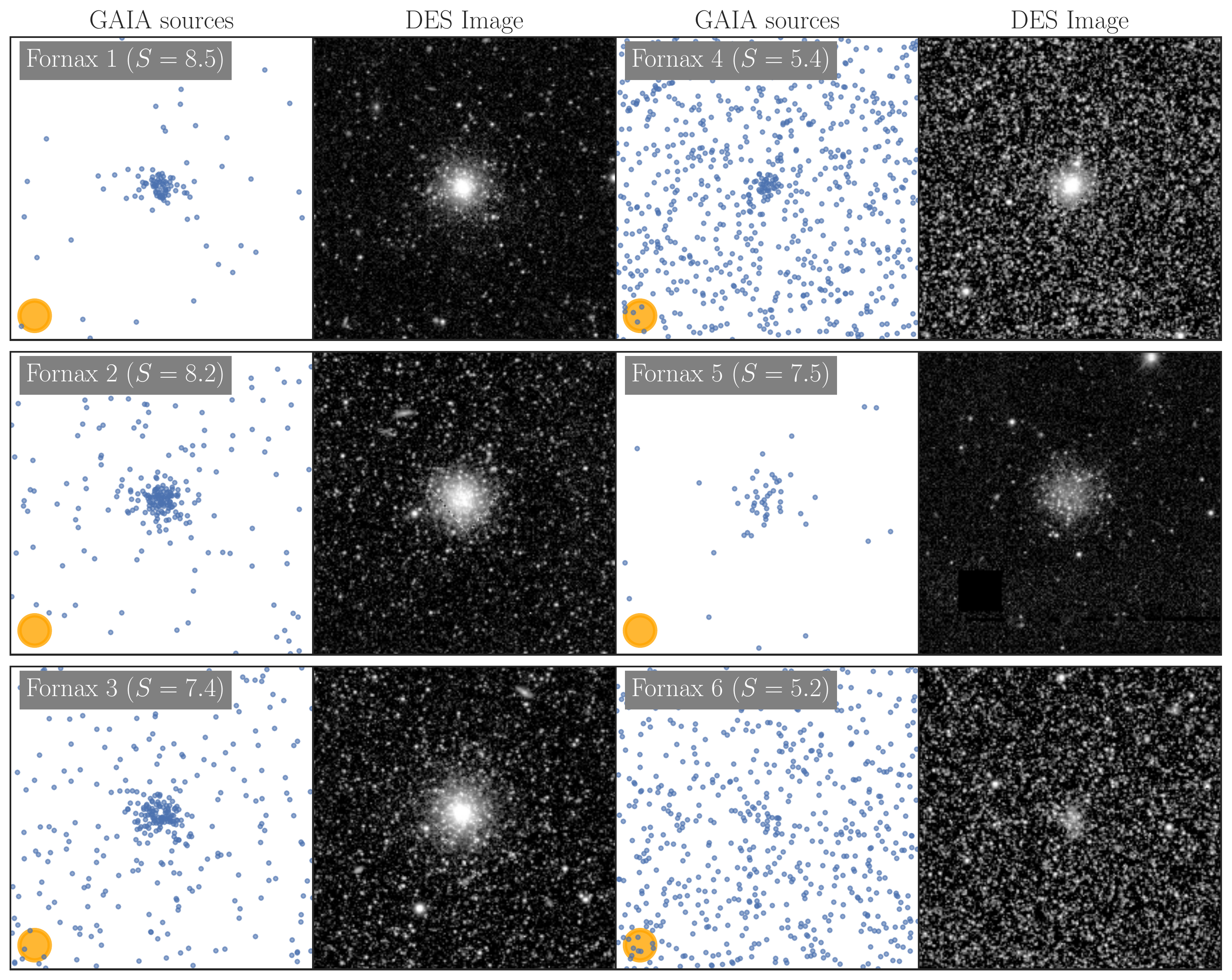}
    \caption{The source maps and the images of the six GCs of Fornax. 
    \textbf{Left} and \textbf{Middle-right} panels: the stellar distributions of \gaia sources centered at each over-density satisfying the detection criteria. The legends show the names of GCs and their significance values $S$. The yellow circles illustrate the inner kernel size of 10~pc. The dimension of each panel is $100 \times 100$~pc$^2$. 
    \textbf{Middle-left} and \textbf{Right} panels: the corresponding images from DES DR1 made with the \code{HiPS}.} 
    \label{fig:known_images}
\end{figure*}


The objective of the paper is to search for possibly missing GCs around the MW satellites by identifying stellar over-densities with the searching algorithm described in Section~\ref{subsec:kde}. 
The list of dwarf galaxies considered in this paper was created by selecting dwarf galaxies within the distance of 450~kpc from the Galactic Center with the exception of the LMC, the SMC, and the Sagittarius dwarf.  
The dwarf list in Table~\ref{tab:target-dwarfs} summarizes all 55 targeted dwarfs investigated in the paper and their properties. 
The reason to exclude the three most massive satellites of the MW is that their relatively large sizes will lead to a huge portion of the sky to be searched, which conflicts with our goal of conducting a targeted search. 
In the construction of the dwarf galaxy list, we use the data from the  \citet{2012AJ....144....4M} compilation and include some of the recent discoveries: Antlia~2 \citep{2019MNRAS.488.2743T}, Aquarius~2 \citep{2016MNRAS.463..712T}, Bootes~3 \citep{2018AA...620A.155M}, Carina~2 \citep{2018MNRAS.475.5085T}, Carina~3 \citep{2018MNRAS.475.5085T}, Cetus~3 \citep{2018PASJ...70S..18H}, Crater~2 \citep{2016MNRAS.459.2370T}, 
and Virgo~I \citep{Homma_2016}.

For each targeted dwarf, we search the area within the radius of ${\rm min} \{ 8^{\circ}, R_{\rm vir} \}$, where $R_{\rm vir}$ is the virial radius of a $10^9 M_{\sun}$ halo \citep{2007ApJ...667L..53W} (at the distance of 100~kpc this corresponds to $10^{\circ}$). 
We choose the inner kernel sizes of $\sigma_1 = 3$, 5, and 10~pc which covers the range of physical sizes of a typical GC \citep{doi:10.1146/annurev.astro.44.051905.092441}. 
We run the searching algorithm for each inner kernel size on the \gaia sources after the selections of Equation~\ref{eq:pmcut} if the dwarf has known measured PM (see Table~\ref{tab:target-dwarfs}), Equation~\ref{eq:g_cuts}, and Equation~\ref{eq:astro_noise_cuts}. 

To balance the completeness of search with the number of false positives, we define two thresholds for identifying possible candidates: a significance threshold $S > 5$ and the limit of the number of stars inside  the inner kernel $\Sigma_{\rm in} > 10$. 
For $S > 5$, as the z-score of the standard normal distribution, its false alarm probability is of the order of $10^{-7}$.\footnote{
    $\int_{5}^{\infty} \frac{1}{\sqrt{2\pi}} e^{-0.5 z^2} dz \sim 10^{-7}$
} 
Assuming a targeted dwarf at the distance of 100~kpc with a searching radius of $8^{\circ}$, the total number of spatial pixels is around the order of $10^8$.\footnote{
    The searching radius of $8^{\circ}$ corresponds to $\sim 10^4$~pc at the distance of 100~kpc so the searching area is $\sim 10^8$~pc$^2$. With the spatial resolution $\sim 1$~pc$^2$, the total number of pixels is then $\sim 10^8$.
} 
With the false alarm probability $\sim 10^{-7}$ on the targeted area of $\sim 10^8$ pixels, the number of expected false positives is around the order of 10. 
Moreover, we apply the other threshold, $\SigmaIn > 10$, to prevent a large number of false positives for the pixels with very low background number density. 
For example in Figure~\ref{fig:method_ex_Fornax}, it is noticeable that the significance can easily be large in the area with very sparse stellar density even if only a handful of stars are detected in the inner kernel. 
These pixels typically have $\SigmaOut < 1$ where the significance estimator breaks down due to the very low rate parameter of Poisson. 
Hence by applying $\Sigma_{\rm in} > 10$, we effectively increase the threshold on $S$ for pixels with $\SigmaOut < 1$, e.g. the threshold is $S=5.6$ for $\SigmaOut = 1$ and $S=8.9$ for $\SigmaOut = 0.1$. 
This avoids the detection of false-positive peaks due to Poisson noise in the $\Sigma_{\rm bg}$ estimates, binary stars, or unresolved galaxies in \gaia that are expected to show more clustering than stars. 
Particularly for binary star systems or unresolved galaxies, the pairs of them are much more likely to occur because they are more correlated; thus they are likely to reach 5 significance and cause false positives. 

After running the searching algorithm on all 55 targeted dwarfs, we identify eleven stellar over-density candidates, based on the highest detected significance of each candidate if it is detected multiple times with different searching parameters. 
Cross-matched with the \code{simbad} database \citep{2000A&AS..143....9W}, all eleven candidates are known objects. 
Nine of them are known GCs: Fornax GC 1\hyrange5 \citep{1938Natur.142..715S, 1961AJ.....66...83H}, Fornax GC 6 \citep{1939PNAS...25..565S, 1981AJ.....86..357V, 1994AJ....108.1648D, 1998PASP..110..533S, 2019ApJ...875L..13W}, Messier 75 \citep{1927BHarO.849...11S}, NGC 5466 \citep{1927BHarO.849...11S}, and Palomar 3 \citep{1955PASP...67...27W}. 
The other two of them are known galaxies: the Leo~I dwarf spheroidal galaxy \citep{1950PASP...62..118H} and the Sextans~A dwarf irregular galaxy \citep{PhysRev.61.489}. 
We remark that Leo~I is found when searching for overdensities near Segue~I, as they are close to each other in the sky.
Table~\ref{tab:known_gcs_found} summarizes the eleven known objects and their detected positions (RA and Dec), significance values ($S$), and inner kernel sizes ($\sigma_1$). 
Figure~\ref{fig:known_images} shows the stellar distribution of \gaia sources for the six GCs of Fornax and the corresponding images from DES DR1 \citep{2018ApJS..239...18A} made with the \code{HiPS} \citep[Hierarchical Progressive Surveys,][]{2015A&A...578A.114F}. 
The yellow circles show the inner kernel of 10~pc (note that it happens to be that all the significance values with 10~pc are greater than with 3 or 5~pc in our detection of the nine GCs). 
Most of those known GCs are detected with the strong significance of $S > 7$ except for Fornax GC 6 with $S = 5.2$, which emphasizes that our algorithm can detect GCs from the regions of high stellar density such as Fornax GC 6. 
Figure~\ref{fig:known_images} further indicates that the significance values are reasonable: bright GCs located at low-density areas (e.g. Fornax GC 1 and 2) have high significance ($S > 8$) and faint GCs located at high-density areas (e.g. Fornax GC 6) have low significance ($S \sim 5$). 
However, we are aware of missing the ultra-faint GC in the Eridanus~2 in our detection, which we will further discuss later in Section~\ref{subsec:completeness}. 



\section{Discussion}
\label{sec:discussion}

\begin{figure*}
    \includegraphics[width=2\columnwidth]{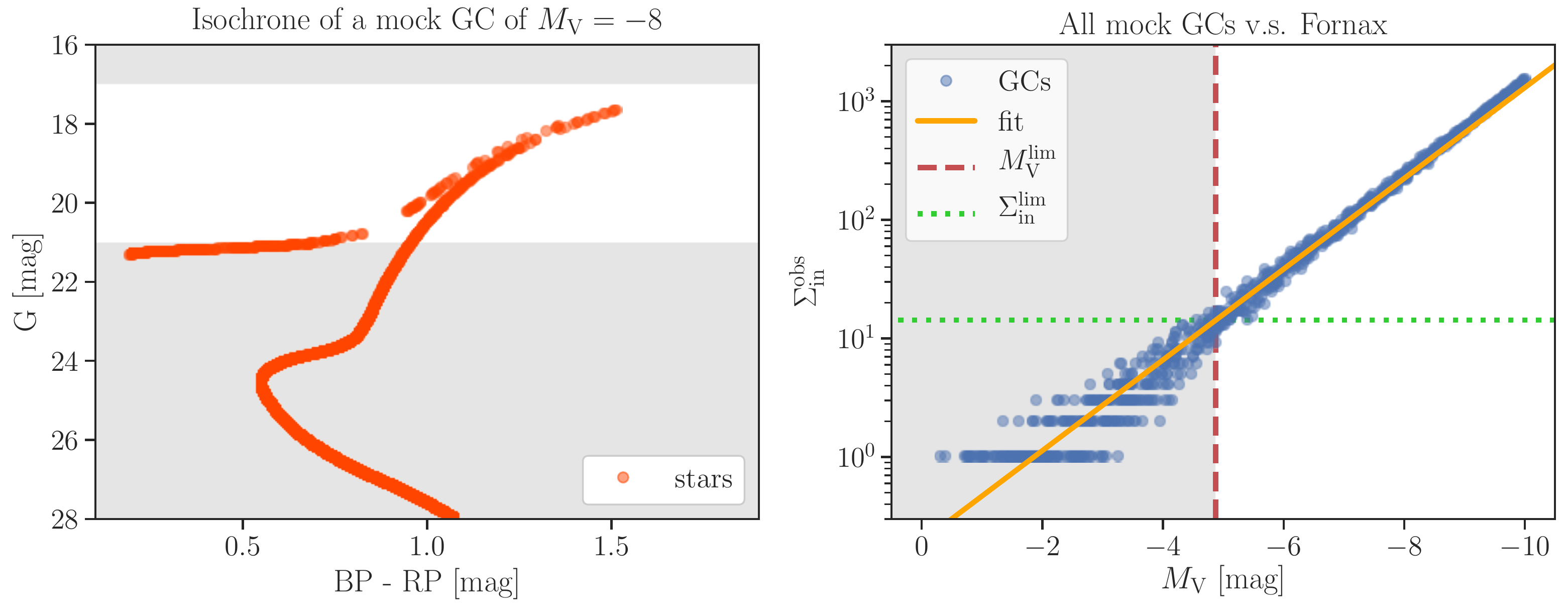}
    \caption{\textbf{Left}: the isochrone of a single mock GC of $\Mv = -8$ at the distance of the Fornax dwarf spheroidal. The stars in the white area are observable within our \gaia G-band cut. \textbf{Right}: the numbers of observable stars $\SigmaInObs$ versus $\Mv$ of all 1000 mock GCs for Fornax. The green dashed line shows the threshold number of stars $\SigmaInLim$ to reach 5 significance according to the maximum background estimate of Fornax. The yellow line is the linear best fit and the red dashed line is the detection limit $\MvLim$ derived based on the best fit and $\SigmaInLim$. The GCs in the white area are detectable.} 
    \label{fig:NobsFitFornax}
\end{figure*}

\begin{figure*}
    \includegraphics[width=2\columnwidth]{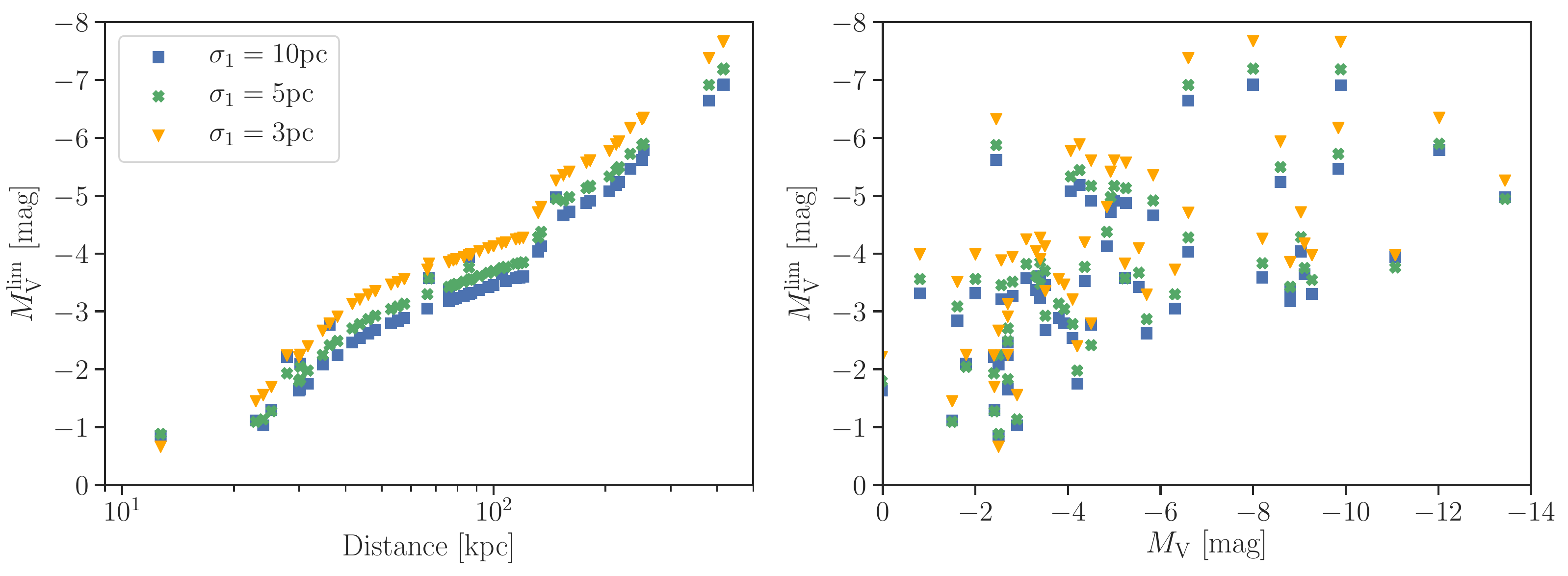}
    \caption{The detection limit $\MvLim$ of all targeted dwarfs with the different inner kernels $\sigma_1=3$, 5, and 10~pc. \textbf{Left}: $\MvLim$ versus the distance of the dwarfs. \textbf{Right}: $\MvLim$ versus the $\Mv$ of the dwarfs.} 
    \label{fig:Mv_lim}
\end{figure*}

\subsection{Detection limit in V-band magnitude}
\label{subsec:detection_limit}

In this section, we will demonstrate how we carry out the detection limit in V-band magnitude $\MvLim$ of the search for each targeted dwarf, which indicates that GCs brighter than $\MvLim$ are detectable in our search. 
To do so, we generate 1000 mock GCs with luminosity in the range of $-10 < \Mv < 0$ assuming the age~$=12$~Gyr and $\rm [Fe/H] = -2$ of the stellar populations. 
Sampling the stars of each GC population according to the log-normal initial mass function in \citet{10.1007/978-1-4020-3407-7_5}, we interpolate the isochrone based on the \textsc{parsec} isochrone \citep{2012MNRAS.427..127B}, then utilizing the isochrones of all the mock GC stellar populations to carry out the detection limit for each targeted dwarf as follows. 

Given a targeted dwarf, to compute the detection limit $\MvLim$, we first calculate the number of observable stars of each mock GC satisfying the G-band selection by counting the number of stars within $17 < \G <21$ according to its isochrone at the distance of the dwarf. 
Based on the number of observable stars, we compute the number of stars of each GC within the inner kernel size $\sigma_1$ as
\begin{equation}
\label{frac_star_mock_gc}
    \SigmaInObs = f \left( \sigma_1;\, r_{\rm h} = 3 {\rm pc} \right) \times \left( {\rm total\ number\ of\ observable\ stars} \right),
\end{equation}
where $f \left( \sigma_1;\, r_{\rm h} \right) = \frac{\sigma_1^2}{\sigma_1^2 + r_{\rm h}^2}$ is the fraction of the number of stars within the radius of $\sigma_1$ according to the Plummer model of 2D surface density profile of a GC with a half-light radius $r_{\rm h} = 3 {\rm pc}$ \citep{1911MNRAS..71..460P}. 
With $\SigmaInObs$ of all the mock GCs at hand, we then use a linear best fit to describe the relation between $\log_{10} \left( \SigmaInObs \right)$ and $\Mv$ of the GCs. 
According to the maximum background estimate of the given dwarf, we know the threshold number of stars $\SigmaInLim$ to be observed to reach 5 significance. 
By comparing $\SigmaInLim$ to the best fit, we can obtain the detection limit $\MvLim$ for the given targeted dwarf.  

We take the Fornax dwarf as an example of the procedure of injection of mock GCs. 
In the left panel of Figure~\ref{fig:NobsFitFornax}, we show the isochrone of a single mock GC of $\Mv = -8$ at the distance of Fornax and the stars in the white area are observable within our \gaia G-band cut. 
By counting the number of stars satisfying $17 < G <21$ corrected by the fraction of stars located within the inner kernels, we know the number of observable stars $\SigmaInObs$ for the given mock GC. 
Applying the calculation of $\SigmaInObs$ for each mock GC, we show the relation between $\SigmaInObs$ and $\Mv$ for all the mock GCs in the right panel of Figure~\ref{fig:NobsFitFornax}. 
The green dashed line shows the threshold number of stars $\SigmaInLim$ to reach $S=5$ according to the maximum background estimate of Fornax; that is, the GCs above the green dashed line are expected to be detectable. 
Fitting the relation between $\log_{10} \left( \SigmaInObs \right)$ and $\Mv$ with a linear best fit as shown in the yellow line, we solve the detection limit $\MvLim$ by finding the value of $\Mv$ satisfying the fit at the value of $\SigmaInLim$ (the green dashed line). 
The red dashed line indicates the derived $\MvLim$ and the GCs brighter than $\MvLim$ in the white area are thus detectable in our search. 
It is worth noting that the \gaia magnitude limit is brighter than $\G = 21$ in some areas of the sky, which will decrease $\SigmaInObs$ if it happens in our targeted area, resulting in a brighter $\MvLim$. 

Repeating the same calculation of $\MvLim$ for all the targeted dwarfs, we obtain the detection limits of the dwarfs and show the comparison of the derived $\MvLim$ to the distances and the luminosities of the dwarfs in Figure~\ref{fig:Mv_lim}. 
In the left panel, there is an obvious trend that the $\MvLim$ are fainter for the dwarfs that are closer because the injected $\SigmaInObs$ of the GCs for these dwarfs with small distance modulus is typically larger than that of the dwarfs with large distance modulus.
On the other hand in the right panel, the relation between $\MvLim$ and $\Mv$ of the dwarfs is more scattered yet there is a slight trend of fainter $\MvLim$ for the fainter dwarfs. 
This is likely because the faint dwarfs, compared to the bright ones, tend to have less-crowded stellar distributions and hence lower thresholds $\SigmaInLim$ to reach 5 significance. 
To sum up, the faint $\MvLim$ for the close dwarfs or the faint dwarfs is reasonable because the ability of dwarfs to hide GCs from our detection is intuitively weaker for the dwarfs that are closer or fainter. 
It is also worth noting that most of the time $\MvLim$ with $\sigma_1 = 10$~pc is the faintest, $\MvLim$ with $\sigma_1 = 5$~pc is the intermediate, and $\MvLim$ with $\sigma_1 = 3$~pc is the brightest mainly because the fractions of stars observed within the inner kernels are around 0.9, 0.7 and 0.5 for $\sigma_1 = 10$, 5, and 3~pc respectively according to the Plummer model. 
That is, the low $\SigmaInObs$ due to the small fraction for small $\sigma_1$ makes the faint GCs less likely to meet 5 significance, thus resulting in a bright $\MvLim$. 

\begin{figure*}
    \includegraphics[width=2\columnwidth]{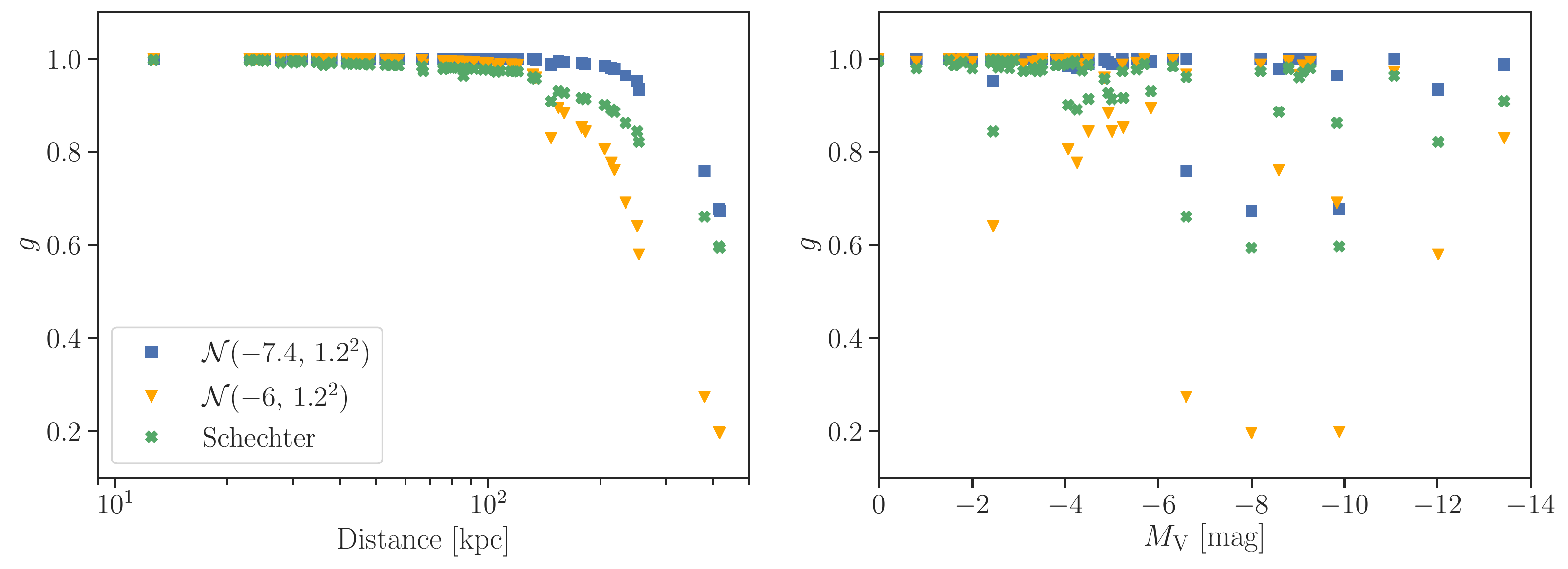}
    \caption{The completeness $g$ of the GC search for all targeted dwarfs with three GCLFs: the Gaussian of $\Gaussian{-7.4}{1.2}$ and $\Gaussian{-6}{1.2}$ and the evolved Schechter in \citet{2007ApJS..171..101J}. \textbf{Left}: Completeness versus the distance of the dwarfs. \textbf{Right}: Completeness versus dwarf galaxy luminosity.} 
    \label{fig:completeness}
\end{figure*}

\subsection{Completeness of the search}
\label{subsec:completeness}

With the limiting magnitudes of GC detection at hand, we can calculate the completeness of the search according to the typical GC luminosity function (GCLF). 
In this section, we will calculate the completeness factor $g$ with three different GCLFs: (a) the typical MW GCLF in \citet{2001stcl.conf..223H}: a Gaussian distribution with a peak at $\Mv = -7.4$ and a standard deviation of 1.2, $\Gaussian{-7.4}{1.2}$, (b) the evolved Schechter function in \citet{2007ApJS..171..101J} with a peak at $\Mv \sim -7.4$, and (c) a presumed Gaussian distribution with a peak at $\Mv = -6$ and a standard deviation of 1.2, $\Gaussian{-6}{1.2}$. 
We calculate $g$ by evaluating the cumulative distribution functions of those GCLFs at $\MvLim$ based on the search with $\sigma_1 = 10$~pc thanks to its better detecting sensitivity compared to $\sigma_1 = 3$ and 5~pc (all the detected objects with the highest significance are detected with $\sigma_1 = 10$~pc in Section~\ref{sec:result}).

We begin with the GCLF in (a); in the MW, the GCLF is approximately a Gaussian distribution of $\Gaussian{-7.4}{1.2}$ \citep{2001stcl.conf..223H}.
With this MW GCLF, we compute the completeness factor $g$ and show them in the blue points in Figure~\ref{fig:completeness}.
The completeness of the search is higher than 90 percent for most of the dwarfs and around 70 percent for the lowest three, Eridanus~2, Leo~T, and Phoenix. 
This high completeness is a consequence of $\MvLim > -7$ for all the dwarfs; that is, the detection limits are fainter than the peak magnitude of the MW GCLF. 
Besides, as a result of the trend of brighter $\MvLim$ for the farther targeted dwarfs in the left panel of Figure~\ref{fig:Mv_lim}, the completeness gets lower for the dwarfs that are more distant. 
In addition, \citet{2010ApJ...717..603V} described that the dispersion of GCLF can be as
small as 0.5 for small dwarfs. 
Calculating the completeness with this GCLF, we find that the result is almost the same as that of the MW GCLF.

Compared to the Gaussian MW GCLF peaking at $\Mv = -7.4$, the evolved Schechter function with a similar peak magnitude proposed in \citet{2007ApJS..171..101J} can describe the GCLF well too, particularly taking good care of the low-mass faint GCs. 
We compute the completeness factor $g$ with this GCLF as shown in the green points in Figure~\ref{fig:completeness}, finding that the difference in $g$ with this GCLF from the traditional Gaussian is less than 5\hyrange10 percent lower. 
The reason for the larger difference ($\sim 10$ percent) in $g$ of the two GCLFs for the targeted dwarfs that are more distant than 100~kpc is that the probability density of the evolved Schechter function is higher than that of the Gaussian MW GCLF in the faint end. 
Thus as these dwarfs have brighter $\MvLim$ than the close dwarfs, their cumulative distribution functions at $\MvLim$ of the evolved Schechter GCLF are lower than that of the Gaussian MW GCLF. 
On the other hand, for the dwarfs that are closer than 100~kpc, $\MvLim$ is much fainter than the peaks of the two GCLFs so the corresponding $g$ approaches 1 for both GCLFs.

So far, we have assumed the GC population for all the dwarfs follows the GCLFs based on the results from bright galaxies, the Gaussian in \citet{2001stcl.conf..223H} and the evolved Schechter in \citet{2007ApJS..171..101J}. 
These two GCLFs have similar peaks but different shapes: the evolved Schechter one extends more toward the faint end to account for faint GCs (see the black curves in Figure~\ref{fig:GCLFs}). 
However, these GCLFs might not hold in the faint host galaxies such as the faint satellites of the MW since there has been no reason for them being universal. 
Especially some of the dwarfs investigated in the paper are even fainter than the peak magnitude of these GCLFs, whether such systems may host GCs that are brighter than the dwarfs themselves is unclear, and is probably unlikely. 
Despite the lack of robust constraints on this, \citet{2006AJ....131..304V} has pointed out that the peak of GCLF can be at $\Mv = -5$ for faint galaxies. 
Moreover, the peak magnitude of GCLFs for different galaxies can vary in the range of $-7 < \Mv < -5$ \citep[see][Table~1 in particular]{2003LNP...635..281R}. 
Therefore, we look at the known GC populations of the MW, NGC 6822, Sagittarius, Fornax, and Eridanus~2 in Appendix~\ref{sec:appendix_GCLF} and decide to consider the peak of GCLF at $\Mv = -6$ based on Figure~\ref{fig:GCLFs} to calculate the completeness again. 
The orange points in Figure~\ref{fig:completeness} show the completeness $g$ computed with the GCLF $\Gaussian{-6}{1.2}$. 
As this GCLF peaks at the fainter magnitude than the other two GCLFs, $g$ hardly changes for close dwarfs with much fainter $\MvLim$ than the peak of GCLF at $\Mv = -6$ whereas $g$ drops for the ones that are more distant than 100~kpc with small $\MvLim$, e.g. $g =$~20\hyrange30 percent for Eridanus~2, Leo~T, and Phoenix.

Section~\ref{sec:result} has mentioned that the ultra-faint GC with the luminosity of $\Mv = -3.5$ \citep{Koposov_2015,2016ApJ...824L..14C} in the Eridanus~2 is missing in our detection. 
This is mainly because the luminosity of this GC is much fainter than the detection limit $\MvLim \sim -6.5$ for the Eridanus~2 in the search. 
Hosting the ultra-faint GC of $\Mv = -3.5$ and having the luminosity of $\Mv = -6.6$ close to the peak magnitude of the MW GCLF, the Eridanus~2 is likely to have a GCLF peaking at a fainter magnitude than $\Mv = -7.4$. 
As shown in Figure~\ref{fig:completeness}, the completeness $g$ for the Eridanus~2 is 75 percent with the Gaussian MW GCLF and 65 percent with the evolved Schechter GCLF. 
When we shift the peak of GCLF to $\Mv = -6$, the completeness factor drops to only 30 percent for the Eridanus~2, which further explains the existence of the ultra-faint GC in the Eridanus~2 while it is missing in our search. 

\begin{figure*}
    \includegraphics[width=2\columnwidth]{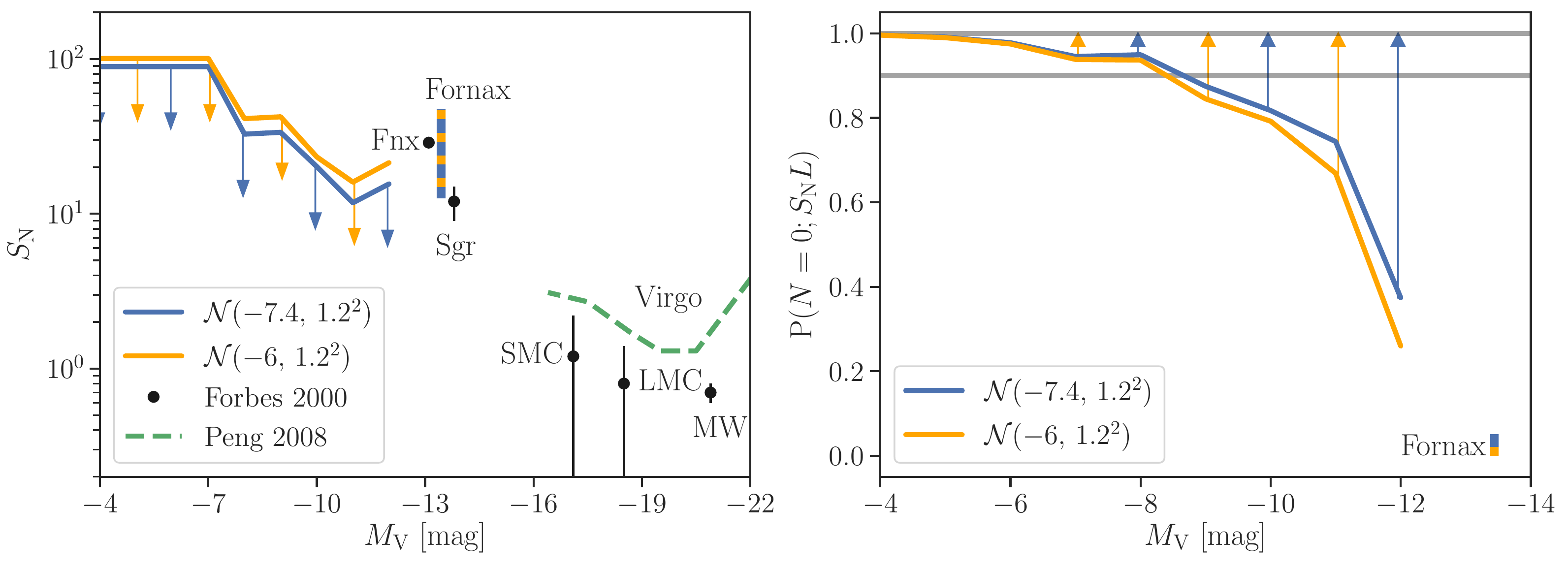}
    \caption{\textbf{Left}: The 90 percent credible intervals on $\SN$ versus $\Mv$ of the dwarfs with two different GCLFs: double-sided intervals for Fornax and one-sided upper bounds for the others. The black data points are $\SN$ of the MW, LMC, SMC, Sagittarius (Sgr), and Fornax (Fnx) in \citet{2000A&A...358..471F}. The green dashed curve is the mean trend curve of the $\SN$ for 100 galaxies in the Virgo Cluster in \citet{2008ApJ...681..197P}. 
    \textbf{Right}: The probability of hosting no GC for a galaxy with luminosity $L$ and specific frequency $\SN$, $P(N=0; \SN L)$. The 90 percent credible intervals on $\SN$ is used to derive the range of $P(N=0; \SN L)$. The two greys lines indicate $P(N=0; \SN L) = 0.9$ and $P(N=0; \SN L) = 1$. 
    } 
    \label{fig:specific_frequency}
\end{figure*}

\subsection{Specific frequency of the globular clusters}
\label{subsec:specific_frequency}

The specific frequency of GCs is a common quantity to indicate the richness of GC system for a galaxy, first formulated as $\SN = \Ngc \times 10^{0.4 \, \left( M_{\rm V, gal} + 15 \right)}$ where $\Ngc$ is the total number of GCs in a host galaxy and $M_{\rm V, gal}$ is the absolute magnitude of the host galaxy \citep{1981AJ.....86.1627H}. 
With $\Lgal \equiv 10^{-0.4 \, \left( M_{\rm V, gal} + 15 \right)}$ defined as the galactic V-band luminosity normalized to $\Mv = -15$, $\SN = \Ngc / \Lgal $ then indicates the number of GCs per unit normalized luminosity. 
When the galaxy luminosity and the number of clusters are large, simply taking a ratio between the number and luminosity makes sense; however, a more statistical approach is required for dwarf galaxies.

Here, we define $\SN$ as the specific frequency for a group of galaxies. 
In that case, the observed number of clusters for each galaxy in a group will be Poisson distributed:
\begin{equation}
\label{eq:poisson_1dwarf_totgc}
    \Ngc \sim {\rm Poisson} \left( \SN \Lgal \right)
\end{equation}
where $\Lgal$ is the luminosity of the galaxy and $\Ngc$ is the random variable describing the number of clusters in this galaxy.
Assuming that our samples of GCs are incomplete with different completeness correction $g$ for each dwarf, we can update the model to include incompleteness as
\begin{equation}
\label{eq:poisson_1dwarf}
    \Ngc \sim {\rm Poisson} \left( \SN g \Lgal \right). 
\end{equation}

Among the nine objects that we identify in our search in Section~\ref{sec:result}, only the six GCs found around the Fornax dwarf are associated with the parent dwarf galaxy. 
That is, the dwarfs targeted in the paper except for Fornax have no associated GCs detected around them. 
Due to the lack of associated GCs and the fact that most of the dwarfs are much fainter than Fornax, the formal $\SN$ is hence expected to be zero with large upper bounds. 
To properly take into account the non-detections and to still be able to constrain the specific frequency of the dwarf population, we assume that $\SN$ is constant for the dwarfs with similar luminosities and will provide upper bounds on $\SN$ for the dwarf population as a whole.

Assuming that we look at $m$ dwarfs as a group at once, we know the luminosity $L_i$ and the completeness $g_i$ for the $i^{\rm th}$ dwarf, where $ 1 \leq i \leq m $. 
The total expected number of observed GCs in this group of $m$ dwarfs is the sum of the expected number of GCs in each dwarf. 
Defining $L \equiv \sum_{i=1}^m L_i g_i$ and with the constant specific frequency $\SN$ shared among the $m$ dwarfs, we can write down the total expected number of GCs as
\begin{equation}
\label{eq:expected_Ngc_sum}
    \sum_{i=1}^m \SN g_i L_i  = \SN \sum_{i=1}^m g_i L_i \equiv \SN L.
\end{equation}
Together with the definition of the total number of observed GCs of the $m$ dwarfs as $N =\sum_{i=1}^m N_i$ where $N_i$ is the number of observed GCs of the $i^{\rm th}$ dwarf from our detection. We model $N$ similarly to Equation~\ref{eq:poisson_1dwarf} as $N \sim {\rm Poisson} \left( \SN L \right)$ and therefore the likelihood function $P(N \mid \SN) \propto \SN^N e^{-\SN L}$. 
Using the Jeffreys prior $\SN^{-1/2}$ as the distribution of the parameter $\SN$, we have the posterior distribution
\begin{equation}
\label{eq:posterior1}
    P(\SN \mid N)  
        \propto  P(\SN) P(N \mid \SN)  
        \propto  \SN^{N - \frac{1}{2}} e^{-\SN L}.
\end{equation}
This is a Gamma distribution; that is, $\SN \sim {\rm Gamma} \left( N + \frac{1}{2}, L \right)$.

With the posterior in Equation~\ref{eq:posterior1}, we construct the 90 percent credible intervals on the parameter $\SN$ with the Gaussian MW GCLF for the dwarfs as shown in the blue curve in the left panel of Figure~\ref{fig:specific_frequency}.  
Also, we show the $\SN$ of the MW and its four most luminous satellites (LMC, SMC, Sagittarius, and Fornax) based on \citet{2000A&A...358..471F} and the mean trend curve of $\SN$ of 100 galaxies in the Virgo Cluster from \citet{2008ApJ...681..197P}. 
Separating the Fornax dwarf from the others due to its richness of GCs, we first calculate its double-sided credible interval on the specific frequency of $12 < \SN < 47$. 
For the other dwarfs with no discovered GCs, we bin the ones brighter than $\Mv = -7$ with a window width of $2$ mag and look at the others all at once, where the value of $\Mv = -7$ is chosen as it is close to the peak magnitude of the GCLF. 
For the dwarfs in each bin, we obtain the one-sided credible intervals as the upper bounds of the specific frequency: $\SN < 20$ for the dwarfs with $-12 < \Mv < -10$, $\SN < 30$ for the dwarfs with $-10 < \Mv < -7$, and $\SN < 90$ for the dwarfs with $\Mv > -7$. 
Similarly, we also construct the credible intervals on $\SN$ with the evolved Schechter GCLF for the dwarfs, finding a similar result as with the Gaussian MW GCLF. 
The difference in $\SN$ with the two GCLFs is less than 5\hyrange10 percent so we only show the one with $\Gaussian{-7.4}{1.2}$ in Figure~\ref{fig:specific_frequency}.

The reason for grouping the dwarfs fainter than $\Mv = -7$ is that they are in general faint so the expected number of GCs is much smaller than one, which makes them not very informative. 
Besides, the posterior becomes more prior-dependent for the fainter dwarfs as well. 
Thus, finding no GCs for the dwarfs in the brighter $\Mv$ bins constrains the upper bounds stronger than in the fainter bins. 
Especially at $\Mv < -10$, the relatively low upper bounds indicate that the Fornax dwarf has a relatively higher $\SN$ than the other dwarfs, especially than the ones with $\Mv < -10$.

As mentioned in Section~\ref{subsec:completeness}, the completeness will drop if the GCLF peaks at a fainter $\Mv$ than the typical peak magnitude at $\Mv = -7.4$, which would effectively increase the upper bounds on $\SN$ because the dropping completeness decreases the $L$.  
We, therefore, calculate the credible intervals on $\SN$ again for the dwarfs with the GCLF $\Gaussian{-6}{1.2}$ as the orange curve shows in the left panel of Figure~\ref{fig:specific_frequency}, finding that the upper bounds on $\SN$ with this shifted GCLF (the orange curve) are higher than that with the MW GCLF (the blue curve) as expected. 
This effect is also expected to influence the upper limits more for the fainter dwarfs since the GCLFs are expected to shift more if the host galaxies are fainter; however, the upper limit is already more prior-dependent and less informative on the faint end so this upper limit increasing effect is less influential.

Besides $\SN$, the probability of a galaxy with luminosity $L$ and $\SN$ to host $N$ GCs, $P(N; \SN L)$, is also interesting. 
With the 90 percent credible intervals on $\SN$, we show the range of $P(N=0; \SN L)$ for a galaxy with $L$ based on the model $N \sim {\rm Poisson} \left( \SN L \right)$ in the right panel of Figure~\ref{fig:specific_frequency}, which indicates the probability of a galaxy to host no GCs.
Except for Fornax, the upper limits of $\SN$ result in the lower limits of $P(N=0; \SN L)$.
Based on $P(N=0; \SN L)$, galaxies fainter than $\Mv = -9$ have $P(N=0; \SN L) > 0.9$, which means the probability of these galaxies to have at least one GC is lower than 10 percent. 
Our finding of $P(N=0; \SN L) > 0.9$ for galaxies with $\Mv > -9$ is in agreement with the claims of the lowest galaxy mass of $\sim 10^5 M_{\sun}$ or luminosity $\Mv \sim -9$ to host at least one GC from \citet{2010MNRAS.406.1967G} and \cite{2018MNRAS.481.5592F}.
This may further explain the observation that galaxies less massive than $10^6 M_{\sun}$ tend not to have nuclei \citep{2019ApJ...878...18S} if we assume that the nuclei originate from GCs sunk by dynamical friction to the center. 
Given our constraints on the specific frequency, Eridanus~2 with $\Mv \sim -7$ has $P(N=0; \SN L) \sim 0.95$, which highlights that the GC inside Eridanus~2 is indeed an outlier.


\section{Conclusions}
\label{sec:conclusion}

We have reported the results of the search for possibly hiding GCs around 55 dwarf galaxies within the distance of 450~kpc from the Galactic Center excluding the LMC, SMC, and Sagittarius. 
This was a targeted search around the dwarfs so we excluded those three satellites to avoid a huge portion of the sky to be searched due to their relatively large sizes. 
For each targeted dwarf galaxy, we have investigated the stellar distribution of the sources in \gaia DR2, selected with the magnitude, proper motion, and stellar morphology cuts. 

Using the kernel density estimation and the Poisson statistics of stellar number counts, we have identified eleven stellar density peaks of above 5 significance as possible GC candidates in the targeted area. 
Cross-matching the eleven possible candidates with the \code{simbad} database and existing imaging data, we have found that all of them are known objects: Fornax GC 1\hyrange6, Messier~75, NGC~5466, Palomar~3, Leo I and Sextans A. 
Only the six GCs of Fornax are associated with the parent dwarf galaxy. 

We have calculated the GC detection limit in $\Mv$ for each dwarf using 1000 simulated GCs, finding that $\MvLim > -7$ for all the dwarfs. 
According to the $\MvLim$ of the dwarfs, we have then calculated the completeness of detection with the Gaussian MW GCLF $\Gaussian{-7.4}{1.2}$, the evolved Schechter GCLF peaking at $\MvLim \sim -7.4$, and the assumed Gaussian GCLF $\Gaussian{-6}{1.2}$. 
With the Gaussian MW GCLF and the evolved Schechter GCLF, the completeness of the detection for most of the dwarfs was higher than 90 percent and even that of the lowest three, Eridanus~2, Leo~T, and Phoenix, was around 70 percent. 
With the assumed Gaussian GCLF, the completeness of our search was lower for the dwarfs that are more distant than 100~kpc, such as the Eridanus~2, Leo~T, and Phoenix where it reached 20\hyrange30 percent. 
Using the completeness, we have constructed the 90 percent credible intervals on the GC specific frequency $\SN$ of the MW dwarf galaxies. 
The Fornax dwarf had the credible interval on the specific frequency of $12 < s < 47$, the dwarfs with $-12 < \Mv < -10$ had $\SN < 20$, the dwarfs with $-10 < \Mv < -7$ had $\SN < 30$, and dwarfs with $\Mv > -7$ had non-informative $\SN < 90$. 
Based on these credible intervals on $\SN$, we have derived the probability of galaxies to host GCs given their luminosity, finding that the probability of galaxies fainter than $\Mv = -9$ to possess GCs is lower than 10 percent.

\defcitealias{2018AA...616A..12G}{Gaia Collab. 18b.}
\defcitealias{2018AA...620A.155M}{MH18}
\defcitealias{2012AJ....144....4M}{M12}
\defcitealias{2018MNRAS.475.5085T}{T18}
\defcitealias{2019MNRAS.488.2743T}{T19b}
\defcitealias{2016MNRAS.463..712T}{T16b}
\defcitealias{2016MNRAS.459.2370T}{T16a}
\defcitealias{2018PASJ...70S..18H}{Homma18}
\defcitealias{Hargis_2016}{Hargis16}
\defcitealias{Homma_2016}{Homma16}

\begin{table*}
\caption{The list of properties of the studied dwarf galaxies: the positions ($\alpha$ and $\delta$), the heliocentric distance ($D_{\sun}$), the V-band magnitude ($\Mv$), the proper motions ($\mu_\alpha$ and $\mu_\delta$), the reference (ref.), and the $3~\sigma_\mu = 3 \sqrt{\sigma_{\mu_{\alpha}}^2 + \sigma_{\mu_{\delta}}^2}$ PM uncertainty converted to \kmPerSec at the distance of the dwarf.}
\label{tab:target-dwarfs}
\begin{tabular}{lrrrrlrrlr}
    dwarf               &  $\alpha$         &  $\delta$                         &   $D_{\sun}$          &  $\Mv$            &  ref.$\rm ^a$                     &   $\mu_\alpha$        &  $\mu_\delta$     &  ref.$\rm ^a$                     &           $3~\sigma_{\mu}$                     \\
                        &  [$^\circ$]       &  [$^\circ$]                       &       [kpc]               &  [mag]                            &                             &  [\masPerYr]                      &  [\masPerYr]         &                                   &           [\kmPerSec]                   \\
    \hline
     Antlia 2             &  143.89 & -36.77 &  132.0 &  -9.0 &    \citetalias{2019MNRAS.488.2743T}  &     -0.095 ± 0.018   &    0.058 ± 0.024     &    \citetalias{2019MNRAS.488.2743T}  &      2e+04 \\
     Aquarius 2           &  338.48 &  -9.33 &  107.9 &  -4.4 &    \citetalias{2016MNRAS.463..712T}  &     -0.252 ± 0.526   &    0.011 ± 0.448     &    \citet{2018AA...619A.103F}        &      3e+05 \\
     Bootes I             &  210.02 &  14.50 &   66.4 &  -6.3 &    \citetalias{2012AJ....144....4M}  &     -0.459 ± 0.041   &    -1.064 ± 0.029    &    \citetalias{2018AA...616A..12G}   &      2e+04 \\
     Bootes II            &  209.50 &  12.85 &   41.7 &  -2.7 &    \citetalias{2012AJ....144....4M}  &     -2.686 ± 0.389   &    -0.530 ± 0.287    &    \citet{2018AA...619A.103F}        &      9e+04 \\
     Bootes III           &  209.25 &  26.80 &   46.0 &  -5.7 &    \citetalias{2018AA...620A.155M}   &     -1.210 ± 0.130   &    -0.920 ± 0.170    &    \citetalias{2018AA...620A.155M}   &      5e+04 \\
     Canes Venatici I     &  202.01 &  33.56 &  217.8 &  -8.6 &    \citetalias{2012AJ....144....4M}  &     -0.159 ± 0.094   &    -0.067 ± 0.054    &    \citet{2018AA...619A.103F}        &      1e+05 \\
     Canes Venatici II    &  194.29 &  34.32 &  100.0 &  -4.9 &    \citetalias{2012AJ....144....4M}  &     -0.342 ± 0.232   &    -0.473 ± 0.169    &    \citet{2018AA...619A.103F}        &      1e+05 \\
     Carina               &  100.40 & -50.97 &  105.2 &  -9.1 &    \citetalias{2012AJ....144....4M}  &     0.495 ± 0.015    &    0.143 ± 0.014     &    \citetalias{2018AA...616A..12G}   &      1e+04 \\
     Carina 2             &  114.11 & -58.00 &   36.2 &  -4.5 &    \citetalias{2018MNRAS.475.5085T}  &     1.810 ± 0.080    &    0.140 ± 0.080     &    \citetalias{2018AA...620A.155M}   &      2e+04 \\
     Carina 3             &  114.63 & -57.90 &   27.8 &  -2.4 &    \citetalias{2018MNRAS.475.5085T}  &     3.035 ± 0.120    &    1.558 ± 0.136     &    \citet{2018ApJ...863...89S}       &      2e+04 \\
     Cetus II             &   19.47 & -17.42 &   29.9 &   0.0 &    \citetalias{2012AJ....144....4M}  &                      &                      &                                             &            \\
     Cetus III            &   31.33 &  -4.27 &  251.0 &  -2.4 &    \citetalias{2018PASJ...70S..18H}  &                      &                      &                                             &            \\
     Columba I            &   82.86 & -28.03 &  182.0 &  -4.5 &    \citetalias{2012AJ....144....4M}  &     -0.020 ± 0.240   &    -0.040 ± 0.300    &    \citet{2019ApJ...875...77P}       &      3e+05 \\
     Coma Berenices       &  186.75 &  23.90 &   43.7 &  -4.1 &    \citetalias{2012AJ....144....4M}  &     0.471 ± 0.108    &    -1.716 ± 0.104    &    \citet{2018AA...619A.103F}        &      3e+04 \\
     Crater 2             &  177.31 & -18.41 &  117.5 &  -8.2 &    \citetalias{2016MNRAS.459.2370T}  &     -0.184 ± 0.061   &    -0.106 ± 0.031    &    \citet{2018AA...619A.103F}        &      4e+04 \\
     Draco                &  260.05 &  57.92 &   75.9 &  -8.8 &    \citetalias{2012AJ....144....4M}  &     -0.019 ± 0.009   &    -0.145 ± 0.010    &    \citetalias{2018AA...616A..12G}   &      5e+03 \\
     Draco II             &  238.20 &  64.57 &   24.0 &  -2.9 &    \citetalias{2012AJ....144....4M}  &     1.170 ± 0.297    &    0.871 ± 0.303     &    \citet{2018ApJ...863...89S}       &      5e+04 \\
     Eridanus 2           &   56.09 & -43.53 &  380.2 &  -6.6 &    \citetalias{2012AJ....144....4M}  &     0.160 ± 0.240    &    0.150 ± 0.260     &    \citet{2019ApJ...875...77P}       &      6e+05 \\
     Eridanus 3           &   35.69 & -52.28 &   87.1 &  -2.0 &    \citetalias{2012AJ....144....4M}  &                      &                      &                                             &            \\
     Fornax               &   40.00 & -34.45 &  147.2 & -13.4 &    \citetalias{2012AJ....144....4M}  &     0.376 ± 0.003    &    -0.413 ± 0.003    &    \citetalias{2018AA...616A..12G}   &      3e+03 \\
     Grus I               &  344.18 & -50.16 &  120.2 &  -3.4 &    \citetalias{2012AJ....144....4M}  &     -0.250 ± 0.160   &    -0.470 ± 0.230    &    \citet{2019ApJ...875...77P}       &      2e+05 \\
     Grus II              &  331.02 & -46.44 &   53.0 &  -3.9 &    \citetalias{2012AJ....144....4M}  &     0.430 ± 0.090    &    -1.450 ± 0.110    &    \citet{2019ApJ...875...77P}       &      3e+04 \\
     Hercules             &  247.76 &  12.79 &  131.8 &  -6.6 &    \citetalias{2012AJ....144....4M}  &     -0.297 ± 0.118   &    -0.329 ± 0.094    &    \citet{2018AA...619A.103F}        &      9e+04 \\
     Horologium I         &   43.88 & -54.12 &   79.4 &  -3.4 &    \citetalias{2012AJ....144....4M}  &     0.950 ± 0.070    &    -0.550 ± 0.060    &    \citet{2019ApJ...875...77P}       &      3e+04 \\
     Horologium II        &   49.13 & -50.02 &   78.0 &  -2.6 &    \citetalias{2012AJ....144....4M}  &                      &                      &                                             &            \\
     Hydra II             &  185.43 & -31.99 &  134.3 &  -4.8 &    \citetalias{2012AJ....144....4M}  &     -0.416 ± 0.519   &    0.134 ± 0.422     &    \citet{2018AA...619A.103F}        &      4e+05 \\
     Indus I              &  317.20 & -51.17 &  100.0 &  -3.5 &    \citetalias{2012AJ....144....4M}  &                      &                      &                                             &            \\
     Indus II             &  309.72 & -46.16 &  213.8 &  -4.3 &    \citetalias{2012AJ....144....4M}  &                      &                      &                                             &            \\
     Leo I                &  152.12 &  12.31 &  253.5 & -12.0 &    \citetalias{2012AJ....144....4M}  &     -0.097 ± 0.056   &    -0.091 ± 0.047    &    \citetalias{2018AA...616A..12G}   &      9e+04 \\
     Leo II               &  168.37 &  22.15 &  233.4 &  -9.8 &    \citetalias{2012AJ....144....4M}  &     -0.064 ± 0.057   &    -0.210 ± 0.054    &    \citetalias{2018AA...616A..12G}   &      8e+04 \\
     Leo IV               &  173.24 &  -0.53 &  154.2 &  -5.8 &    \citetalias{2012AJ....144....4M}  &     -0.590 ± 0.531   &    -0.449 ± 0.358    &    \citet{2018AA...619A.103F}        &      5e+05 \\
     Leo V                &  172.79 &   2.22 &  177.8 &  -5.3 &    \citetalias{2012AJ....144....4M}  &     -0.097 ± 0.557   &    -0.628 ± 0.302    &    \citet{2018AA...619A.103F}        &      5e+05 \\
     Leo T                &  143.72 &  17.05 &  416.9 &  -8.0 &    \citetalias{2012AJ....144....4M}  &                      &                      &                                             &            \\
     Pegasus 3            &  336.09 &   5.42 &  205.1 &  -4.1 &    \citetalias{2012AJ....144....4M}  &                      &                      &                                             &            \\
     Phoenix              &   27.78 & -44.44 &  415.0 &  -9.9 &    \citetalias{2012AJ....144....4M}  &     0.079 ± 0.099    &    -0.049 ± 0.120    &    \citet{2018AA...619A.103F}        &      3e+05 \\
     Phoenix 2            &  355.00 & -54.41 &   83.2 &  -2.8 &    \citetalias{2012AJ....144....4M}  &     0.490 ± 0.110    &    -1.030 ± 0.120    &    \citet{2019ApJ...875...77P}       &      6e+04 \\
     Pictoris I           &   70.95 & -50.28 &  114.8 &  -3.1 &    \citetalias{2012AJ....144....4M}  &                      &                      &                                             &            \\
     Pisces II            &  344.63 &   5.95 &  182.0 &  -5.0 &    \citetalias{2012AJ....144....4M}  &     -0.108 ± 0.645   &    -0.586 ± 0.498    &    \citet{2018AA...619A.103F}        &      7e+05 \\
     Reticulum II         &   53.93 & -54.05 &   30.2 &  -2.7 &    \citetalias{2012AJ....144....4M}  &     2.340 ± 0.120    &    -1.310 ± 0.130    &    \citetalias{2018AA...620A.155M}   &      2e+04 \\
     Reticulum III        &   56.36 & -60.45 &   91.6 &  -3.3 &    \citetalias{2012AJ....144....4M}  &     -1.020 ± 0.320   &    -1.230 ± 0.400    &    \citet{2019ApJ...875...77P}       &      2e+05 \\
     Sagittarius II       &  298.17 & -22.07 &   67.0 &  -5.2 &    \citetalias{2012AJ....144....4M}  &     -1.180 ± 0.140   &    -1.140 ± 0.110    &    \citetalias{2018AA...620A.155M}   &      6e+04 \\
     Sculptor             &   15.04 & -33.71 &   85.9 & -11.1 &    \citetalias{2012AJ....144....4M}  &     0.082 ± 0.005    &    -0.131 ± 0.004    &    \citetalias{2018AA...616A..12G}   &      3e+03 \\
     Segue I              &  151.77 &  16.08 &   22.9 &  -1.5 &    \citetalias{2012AJ....144....4M}  &     -1.697 ± 0.195   &    -3.501 ± 0.175    &    \citet{2018AA...619A.103F}        &      3e+04 \\
     Segue II             &   34.82 &  20.18 &   34.7 &  -2.5 &    \citetalias{2012AJ....144....4M}  &     1.270 ± 0.110    &    -0.100 ± 0.150    &    \citetalias{2018AA...620A.155M}   &      3e+04 \\
     Sextans I            &  153.26 &  -1.61 &   85.9 &  -9.3 &    \citetalias{2012AJ....144....4M}  &     -0.496 ± 0.025   &    0.077 ± 0.020     &    \citetalias{2018AA...616A..12G}   &      1e+04 \\
     Triangulum II        &   33.32 &  36.18 &   30.2 &  -1.8 &    \citetalias{2012AJ....144....4M}  &     0.651 ± 0.193    &    0.592 ± 0.164     &    \citet{2018ApJ...863...89S}       &      4e+04 \\
     Tucana II            &  342.98 & -58.57 &   57.5 &  -3.8 &    \citetalias{2012AJ....144....4M}  &     0.910 ± 0.060    &   -1.160 ± 0.080     &    \citet{2019ApJ...875...77P}       &      3e+04 \\
     Tucana III           &  359.15 & -59.60 &   25.2 &  -2.4 &    \citetalias{2012AJ....144....4M}  &     -0.030 ± 0.040   &    -1.650 ± 0.040    &    \citet{2019ApJ...875...77P}       &      7e+03 \\
     Tucana IV            &    0.73 & -60.85 &   48.1 &  -3.5 &    \citetalias{2012AJ....144....4M}  &     0.630 ± 0.250    &    -1.710 ± 0.200    &    \citet{2019ApJ...875...77P}       &      7e+04 \\
     Tucana V             &  354.35 & -63.27 &   55.2 &  -1.6 &    \citetalias{2012AJ....144....4M}  &                      &                      &                                             &            \\
     Ursa Major I         &  158.72 &  51.92 &   96.8 &  -5.5 &    \citetalias{2012AJ....144....4M}  &     -0.659 ± 0.093   &    -0.635 ± 0.131    &    \citet{2018ApJ...863...89S}       &      7e+04 \\
     Ursa Major II        &  132.88 &  63.13 &   31.6 &  -4.2 &    \citetalias{2012AJ....144....4M}  &     1.661 ± 0.053    &    -1.870 ± 0.065    &    \citet{2018ApJ...863...89S}       &      1e+04 \\
     Ursa Minor           &  227.29 &  67.22 &   75.9 &  -8.8 &    \citetalias{2012AJ....144....4M}  &     -0.182 ± 0.010   &    0.074 ± 0.008     &    \citetalias{2018AA...616A..12G}   &      4e+03 \\
     Virgo I              &  180.04 &  -0.68 &   87.0 &  -0.8 &    \citetalias{Homma_2016}           &                      &                      &                                             &            \\
     Willman I            &  162.34 &  51.05 &   38.0 &  -2.7 &    \citetalias{2012AJ....144....4M}  &     0.199 ± 0.187    &    -1.342 ± 0.366    &    \citet{2018AA...619A.103F}        &      7e+04 \\
     \hline
\end{tabular}
\begin{flushleft}
    $\rm ^a$ Some of the citations are abbreviated: 
    \citetalias{2018AA...616A..12G} is for \citet{2018AA...616A..12G}; 
    \citetalias{2018AA...620A.155M} is for \citet{2018AA...620A.155M}; \citetalias{2012AJ....144....4M} is for \citet{2012AJ....144....4M};
    \citetalias{2018MNRAS.475.5085T} is for \citet{2018MNRAS.475.5085T};
    \citetalias{2019MNRAS.488.2743T} is for \citet{2019MNRAS.488.2743T};
    \citetalias{2016MNRAS.463..712T} is for \citet{2016MNRAS.463..712T};
    \citetalias{2016MNRAS.459.2370T} is for \citet{2016MNRAS.459.2370T};
    \citetalias{2018PASJ...70S..18H} is for \citet{2018PASJ...70S..18H};
    \citetalias{Hargis_2016} is for \citet{Hargis_2016};
    \citetalias{Homma_2016} is for \citet{Homma_2016}.
\end{flushleft}
\end{table*}

\section*{Acknowledgements}
We acknowledge the support by NSF grants AST-1813881, AST-1909584, and Heising-Simons Foundation grant 2018-1030. This paper has made use of the Whole Sky Database (wsdb) created by Sergey Koposov and maintained at the Institute of Astronomy, Cambridge with financial support from the Science \& Technology Facilities Council (STFC) and the European Research Council (ERC). This software has made use of the \code{q3c} software \citep{Koposov2006}.

This work presents results from the European Space Agency (ESA) space mission \gaia. \gaia data are being processed by the \gaia Data Processing and Analysis Consortium (DPAC). Funding for the DPAC is provided by national institutions, in particular the institutions participating in the \gaia MultiLateral Agreement (MLA). The \gaia mission website is \href{https://www.cosmos.esa.int/gaia}{https://www.cosmos.esa.int/gaia}. The \gaia archive website is \href{https://archives.esac.esa.int/gaia}{https://archives.esac.esa.int/gaia}.

Software:
\acknowledgeSoftware{numpy},
\acknowledgeSoftware{scipy},
\acknowledgeSoftware{pandas},
\acknowledgeSoftware{matplotlib},
\acknowledgeSoftware{seaborn},
\acknowledgeSoftware{astropy},
\acknowledgeSoftware{imf},
\acknowledgeSoftware{sqlutilpy}.

\section*{Data Availability}

The data underlying this article were derived from sources in the public domain: \href{https://archives.esac.esa.int/gaia}{https://archives.esac.esa.int/gaia}.



\bibliographystyle{mnras}
\bibliography{bib} 


\appendix
\section{GC luminosity functions}
\label{sec:appendix_GCLF}

\begin{figure}
    \includegraphics[width=\columnwidth]{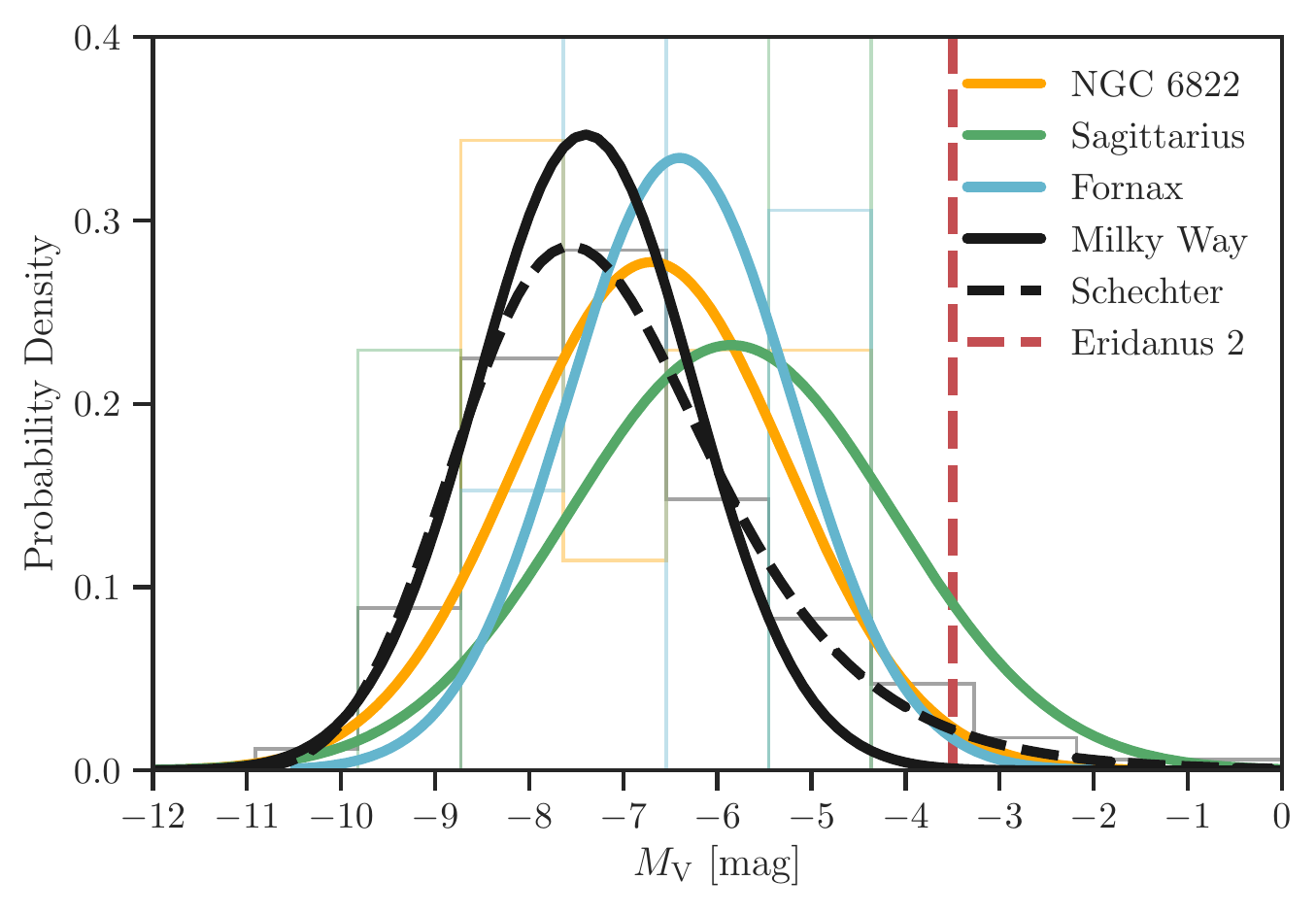}
    \caption{GCLFs of the Milky Way, NGC~6822, Sagittarius, and Fornax. For each galaxy, the solid curve is the Gaussian fit to the histogram of the probability density of the number of GCs in each magnitude bin. The black dashed curve is the evolved Schechter function in \citet{2007ApJS..171..101J}. The red dashed line indicates the ultra-faint GC of the Eridanus~2. 
    } 
    \label{fig:GCLFs}
\end{figure}

In Section~\ref{sec:discussion}, we adopt the Gaussian MW GCLF in \citet{2001stcl.conf..223H} and the evolved Schechter GCLF in \citet{2007ApJS..171..101J} for all the dwarfs to carry out the completeness factor and the specific frequency. 
However, the GCLF may shift toward the faint end for faint dwarfs, e.g. \citet{2003LNP...635..281R,2006AJ....131..304V}.
To investigate this, we show the GCLFs in the histogram with Gaussian probability density distributions of the MW ($\Mv \sim -21$), NGC~6822 ($\Mv \sim -16$), Sagittarius ($\Mv \sim -14 $), and Fornax ($\Mv \sim -13 $) with the solid curves in Figure~\ref{fig:GCLFs}. 
Besides, we also show the evolved Schechter GCLF with the black dashed curve and the ultra-faint GC of the Eridanus~2 ($\Mv \sim -7$) with the red dashed line. 
We collect the GC lists for these galaxies according to \citet{2010arXiv1012.3224H}, \citet{2015MNRAS.452..320V}, \citet{Koposov_2015}, \citet{Vasiliev2019}, or the \code{simbad} database. 
Based on the Gaussian distributions of the GCLFs and the existence of Eridanus~2 GC, there is a possible shift of the GCLF peak toward the faint luminosity for faint galaxies, e.g. the peaks of the dwarf galaxies are closer to $\Mv \sim -6$ as opposed to the peak of the GC distribution in the MW at $\Mv = -7.4$. 

\bsp	
\label{lastpage}
\end{document}